\documentclass[aps,twocolumn,showpacs,nofootinbib]{revtex4-1}

\usepackage{amsmath}
\usepackage{amsfonts,amssymb}
\usepackage{multirow}
\usepackage[dvips]{graphicx}
\usepackage{float}
\usepackage{appendix}
\usepackage{color}
\usepackage{url}
\usepackage{subfigure}
\usepackage{footnote}

\def\beq{\begin{equation}}
\def\eeq{\end{equation}}
\def\bea{\begin{eqnarray}}
\def\eea{\end{eqnarray}}

\def\nl{\nonumber\\}

\def\chic1{\chi_{c1}}

\newcommand{\gsim}{\lower.7ex\hbox{$
\;\stackrel{\textstyle>}{\sim}\;$}}
\newcommand{\lsim}{\lower.7ex\hbox{$
\;\stackrel{\textstyle<}{\sim}\;$}}

\newcommand{\eod}{\end{document}}

\newcommand{\Lambar}{\bar\Lambda}
\newcommand{\Jpsi}{J/\psi}

\RequirePackage{xspace} \allowdisplaybreaks

\begin{document}

\title{CP Asymmetries in Strange Baryon Decays }
\author{I.I.~Bigi$^1$}\email{ibigi@nd.edu}
\author{Xian-Wei Kang$^{2,3}$}\email{kxw198710@126.com}
\author{Hai-Bo Li$^{3,4}$}\email{lihb@ihep.ac.cn}
\affiliation{$^1$Department of Physics, University of Notre Dame du Lac, Notre Dame, IN 46556, USA\\
$^2$Institute of Physics, Academia Sinica, Taipei, Taiwan 115 \\
$^3$Institute of High Energy Physics, Beijing 100049, People's
Republic of China \\
$^4$University of Chinese Academy of Science, Beijing 100049,
People's Republic of China }

\begin{abstract}
 While indirect \& direct CP violations (CPV) had been
established in the decays of strange \& beauty mesons, none have
been done for baryons. There are different "roads" for finding CP
asymmetries in the decays of strange baryons; they are highly
non-trivial ones. The HyperCP Collaboration had probed CPV in the
decays of single $\Xi$ \& $\Lambda$ \cite{HYPER}. We talk about
future lessons from $e^+e^-$ collisions at BESIII/BEPCII: probing
decays of {\em pairs} of strange baryons, namely $\Lambda$, $\Sigma$
\& $\Xi$. Realistic goals are to learn about non-perturbative QCD.
One can hope to find CPV in the decays of strange baryons; one can
also dream to find impact of New Dynamics (ND). We point out that a
new important era starts with the BESIII/BEPCII data accumulated by
the end of 2018. It supports also new ideas to trigger $J/\psi \to
\bar \Lambda \Lambda$ at the LHCb collaboration.
\end{abstract}

\maketitle

\section{The landscape}
\label{History}

CP violation (CPV) has been found in $K_L \to \pi^+\pi^-$ in 1964
\cite{CRONIN} (actually `predicted' by L.B. Okun in 1963
\cite{OKUN})  and tiny {\em direct} CPV in differences $K_L \to
\pi^+\pi^-$ vs. $K_L \to \pi^0\pi^0$ in 1990's by NA31/NA48 and
KTeV. Present analyses give in Particle Data Group 2016
(PDG2016)~\cite{Olive:2016xmw}:
\bea
 |\epsilon_K|_{\rm exp.} &=&
(2.228 \pm 0.011) \cdot 10^{-3} \,, \nl
 {\rm Re}(\epsilon^{\prime}/\epsilon_K)_{\rm exp.} &=& (1.66 \pm 0.23)\cdot
10^{-3} \, ; \eea i.e., CPV in $\Delta S =1$ has been found there
around the scale of $5 \cdot 10^{-6}$. The impact of New Dynamics
(ND) can possibly hide in the uncertainties about {\em direct} CPV.
The Buras team gives for the Standard Model \cite{BURAS1}: \beq {\rm
Re}(\epsilon^{\prime}/\epsilon_K)_{\rm "Buras"} = (0.86 \pm
0.32)\cdot 10^{-3} \; . \eeq Actually, it has been argued by Buras
{\em et al.} for a long time the SM cannot produce the value given
by the data.

Now we have gotten the first result from a LQCD group
\cite{LATTICE1}: \beq {\rm Re}(\epsilon^{\prime}/\epsilon_K)_{\rm
LQCD} = (0.138 \pm 0.515 \pm 0.443)\cdot 10^{-3}. \eeq Obviously we
need more lattice data. So far, it is not clear which lesson we can
find here: does it mean that these data are consistent what the SM
gives us or it is a sign for the impact of ND?

Soon after 1964 it was said to probe CPV in the transitions of strange {\em baryons}.
It is a huge challenge to find it. However, the goal is so important that we should not give up.
Present experimental limits are high about what one can think about even the possible impact of the ND.
In 1998 it had been described in several situations, in particular about CP observables in $B^0(t) \to$ hyperon-antihyperon,
where CPV could be found based on predictions at {\em that} time \cite{RRD98}.

The landscape of fundamental dynamics have changed very much after
the 2nd millennium. Neutrinos are massless in the SM; however
neutrino oscillations have been established. The SM produces true
large CP asymmetries in $\Delta B \neq 0$ transitions, and indeed
the BABAR/Belle Collaborations have found in the beginning of the
3rd millennium with $B^0 \to J/\psi K_S$ \& others. Still no
non-zero values of CP asymmetries have been established in the
decays of baryons in general beyond the huge asymmetry in matter vs.
anti-matter in our Universe (or our `existence'). However, evidence
has been seen by the LHCb experiment in $\Lambda^0_b \to p\pi^-
\pi^+\pi^-$ for regional CPV \cite{LHCbLambda_b}. Obviously we are
talking about direct CPV in $\Delta B =1$. In two-body final states
(FS) of beauty {\em mesons} the usual scale is $\sim 0.1$; for
$\Lambda_b^0$ decays it suggests that a regional one is sizably
enhanced. Can it happen also for strange baryons?

The final states (FS) are mostly produced by two-body ones in the decays of strange baryons. There are two classes of transitions as PDG2016 data give us:
\begin{itemize}
\item
Re-scattering gives sizable impacts, and it is obvious for $\Lambda$
and $\Sigma^{+}$:
\bea
 &&{\rm BR} (\Lambda \to p \pi^-) = 0.639 \pm 0.005 \nl
 &&{\rm BR} (\Lambda \to n \pi^0) = 0.358 \pm 0.005 \nl
 &&{\rm BR} (\Lambda \to p \pi^- \gamma) = (8.4 \pm 1.4) \times 10^{-4}\nl
 &&{\rm BR} (\Sigma^+ \to p \pi^0) = 0.5157 \pm 0.0030 \nl
 &&{\rm BR} (\Sigma^+ \to n \pi^+) = 0.4831\pm 0.0030
\eea
CPT invariance is `usable' telling us about averaged
asymmetries \footnote{The impact of CPT invariance goes well beyond
the same mass \& width as discussed in \cite{CPBOOK}.}. These widths
are basically produced with two hadrons; thus: $\Gamma (\Lambda \to
p \pi^-) + \Gamma (\Lambda \to n \pi^0)$ $\simeq$ $\Gamma (\bar
\Lambda \to \bar p \pi^+) + \Gamma (\bar \Lambda \to \bar n \pi^0)$;
likewise for $\Sigma^+$. Therefore \bea
A_{\rm CP}(\Lambda \to p \pi^-) &\simeq& - A_{\rm CP}(\Lambda \to n \pi^0) \\
A_{\rm CP}(\Sigma^+ \to p \pi^0) &\simeq & - A_{\rm CP}(\Sigma^+ \to
n \pi^+). \eea The goal is to establish CP asymmetries in $\Lambda
\to p \pi^-$ and in $\Sigma^+ \to p \pi^0$. To find it also in
$\Lambda \to n \pi^0$ \& $\Sigma^+ \to n \pi^+$ is nice, but not
important. The situations are very different for the decays of
$\Lambda_b^0$, where the final states are mostly given by many-body
ones, as we know.

When one discusses decays $\Lambda \to p \pi^-$ without the spins of the baryons included, one gets only one observable, namely a number.
The data depend on production asymmetries in $\Lambda \to p \pi^-$ vs. $\bar \Lambda \to \bar p \pi^+$.
That is not a problem for $e^+e^-$ annihilations for the BESIII experiment or for $\bar p p$ collisions; however the situation is quite different for $pp$ ones from LHCb data.
\item
The situations are more complex:  the impact of re-scattering is not
obvious, when one cannot use polarized baryons, as one can see in
the branching ratios:
\begin{eqnarray}
&& {\rm BR} (\Sigma^- \to n\pi^-)=(99.848\pm 0.005) \times
10^{-2}\nl && {\rm BR} (\Sigma^- \to n\pi^-\gamma ) = (4.6 \pm 0.6)
\times 10^{-4}\nl &&{\rm BR} (\Xi^0 \to \Lambda \pi^0) = (99.524 \pm
0.012) \times 10^{-2} \nl &&{\rm BR} (\Xi^0 \to \Lambda \gamma ) =
(1.17 \pm 0.07) \times 10^{-3}\nl && {\rm BR} (\Xi^- \to \Lambda
\pi^-) = (99.887 \pm 0.035) \times 10^{-2} \nl && {\rm BR} (\Xi^-
\to \Sigma^- \gamma ) = (1.27 \pm 0.23) \times 10^{-4}
\end{eqnarray}
\end{itemize}
Probing CPV in strange baryons transitions is a true challenge to be successful. In the end of
the previous millennium we got a predictions
based on the SM \cite{DONOGH86,CPBOOK2}:
\bea
A_{\rm CP} (\Lambda \to p \pi^-) &\sim& (0.05 - 1.2) \cdot 10^{-4} \\
A_{\rm CP} (\Xi^- \to \Lambda \pi^-) &\sim& (0.2 - 3.5) \cdot 10^{-4}
\eea
In the  beginning of this millennium we got a SM prediction by combining $\Lambda$ \& $\Xi$ decays \cite{VALENCIA}:
$-0.5 \cdot 10^{-4} \leq  A_{\Lambda \Xi} \equiv \frac{\alpha_{\Lambda}\alpha_{\Xi} - \alpha_{\bar \Lambda}\alpha_{\bar \Xi} }{\alpha_{\Lambda}\alpha_{\Xi} + \alpha_{\bar \Lambda}\alpha_{\bar \Xi}} \leq
0.5 \cdot 10^{-4}$.

The HyperCP experiment \footnote{The authors of this proposal "HyperCP: Search for CP Violation in Charged-Hyperon Decays" quoted Bigi and Sanda from their recent book {\em CP Violation}:
{\em "We are willing to stake our reputation on the prediction that dedicated and comprehensive studys of
CP violation will reveal the presence of New Physics."}. At least one of the co-authors of this paper agrees.}
had searched for CPV by a 800 GeV proton beam on a Cu target \cite{HYPER}:
\beq
A_{\Lambda \Xi}  = (0.0 \pm 5.1 \pm 4.4 ) \cdot 10^{-4}  \; .
\eeq
It is still not clear, whether the theoretical uncertainties are included correctly.

However, two points might help to reach our goals:
\begin{itemize}
\item
The BESIII collaboration can probe {\em pairs} of strange baryons.
We will discuss that in Sect.\ref{PAIRS}.

\item
Future BESIII analyses will be enhanced, namely $e^+e^- \to J/\psi$
with the unusual narrow resonance as a source of strange baryons
from PDG2016: \bea &&{\rm BR}(J/\psi \to \bar \Lambda \Lambda) =
(1.61 \pm 0.15) \cdot 10^{-3} \nl &&{\rm BR}(J/\psi \to \Lambda \bar
\Lambda \pi^+\pi^-) =  (4.3 \pm 1.0) \cdot 10^{-3}  \nl &&{\rm
BR}(J/\psi \to \Lambda \bar p K^+/\bar \Lambda p K^-) = (0.89 \pm
0.16)\cdot 10^{-3} \nl &&{\rm BR}(J/\psi \to \bar \Sigma ^+
\Sigma^-/\bar \Sigma^-\Sigma^+ ) = ( 1.50 \pm 0.24 ) \cdot 10^{-3}
\nl &&{\rm BR}(J/\psi \to \bar \Xi^0 \Xi^0 ) =( 1.20 \pm 0.24 )
\cdot 10^{-3}\nl &&{\rm BR}(J/\psi \to \bar \Xi^+ \Xi^-/\bar \Xi^-
\Xi^+ ) = (0.86 \pm 0.11   ) \cdot 10^{-3} \eea These rates are
produced by strong forces, and they can be compared with \bea &&{\rm
BR}(J/\psi \to \bar p p ) = (2.120 \pm 0.029) \cdot 10^{-3}\nl
&&{\rm BR}(J/\psi \to \bar p p \pi^+\pi^-   ) = (6.0 \pm 0.5 ) \cdot
10^{-3}\nl && {\rm BR}(J/\psi \to p\bar n \pi^-   )=( 2.12 \pm 0.09
) \cdot 10^{-3}\nl && {\rm BR}(J/\psi \to  \bar n n  ) = (2.09 \pm
0.16 ) \cdot 10^{-3} \eea The final state interactions (FSI) show
impact, although our `community' is so far not able to describe it
quantitatively.

\end{itemize}
We will discuss in some details what we can learn about fundamental dynamics including CP asymmetries. It is not trivial at all; as usual there is a price for a prize.

\section{CP asymmetries in $J/\psi \to$ pair of strange baryons}
\label{PAIRS}

Mostly we discuss the decays of $J/\psi$ to final states with only two strange baryons. We also include
special cases in Sect.\ref{YY}.

First we quickly go back to the history about discrete symmetries,
in particular about "parity conservation, charge-conjugation
invariance, and time-reversal invariance" in the decays of hyperons
\cite{LYHYP}. PDG2016 gives for the decay of $\Lambda \to p \pi^-$ a
T-odd moment $\alpha_{\Lambda} = 0.642 \pm 0.013$, while
$\alpha_{\bar\Lambda} = - 0.71 \pm 0.08$ for $\bar \Lambda \to \bar
p \pi^+$; the dominant data come from the BES Collaboration:
$\alpha_{\bar\Lambda} = - 0.755 \pm 0.083 \pm 0.063$ \cite{BES1}.

Published 2010 data from the BES Collaboration are based on $5.8
\times 10^{7} \; J/\psi$ events~\cite{BES1}; there we are talking
about 10 \% of the accuracy. To reach the level of ${\cal O}(1\%)$
is not trivial. However, we discuss about the level of 0.1 \% and
pointed out that our community will have the data to reach that by
2018. We have to be realistic: this prize cannot be reached easily.
On the other hand, it would be very pessimistic to suggest we will
follow the `road' of the Higgs boson, namely searching for 40 years.

Now it has already
$ \simeq 1.3 \times 10^9$ $J/\psi$ events collected by the BESIII detector.
It expects to get close to $10^{10}$ $J/\psi$ events by the end of 2018~\cite{Li:2016tlt}.
The situations are even more complex, when we discuss BR$(J/\psi \to \bar \Xi \Xi \to [\bar \Lambda \pi][\Lambda \pi])$ in Sect.\ref{XiXiLambda}.

Further competition will arrive from the LHCb collab. by probing
$J/\psi \to \bar \Lambda \Lambda$; the challenge comes by triggering
it \cite{PUNZI}.

\subsection{General statements about first steps}
\label{INTRO}

First we describe the `landscape' and the tools one can use there.
The super narrow vector resonance $J/\psi$ produces a connection of
the spins of a pair of hyperons $Y$ of $\Lambda$, $\Sigma$ \& $\Xi$
and FS  $X$ of $p$, $\Lambda$: \beq J/\psi \to \bar Y Y \to [\bar X
\bar \pi] [X\pi] \eeq One can measure $T$-odd moments: \beq
\alpha^{(X)}_{Y} = \langle  \vec \sigma _{Y} \cdot (\vec \sigma_X
\times \vec \pi_X) \rangle \, , \, \alpha^{(\bar X)}_{\bar Y} =
\langle  \vec \sigma _{\bar Y} \cdot (\vec \sigma_{\bar X} \times
\vec \pi_{\bar X}) \rangle \; , \eeq where $\vec \sigma$ \& $\vec
\pi$ describe the spins \& the momenta of the baryons in the rest
frames of $Y$/$\bar Y$. It is not surprising to find sizable or even
large non-zero values of $T$-odd moments; below we will give an
example from data with actually large values. It shows the impact of
FSI. Non-zero values of $T$-odd moments do {\em not} mean by
themselves we have found $T$ violation -- or CP violation using CPT
invariance. However, one can probe direct CP asymmetries \beq
\langle A^{(X)}_{\rm CP}\rangle =
\frac{\alpha^{(X)}_{Y}+\alpha^{(\bar X)}_{\bar
Y}}{\alpha^{(X)}_{Y}-\alpha^{(\bar X)}_{\bar Y}} \eeq with{\em out}
polarized $Y$ \& $\bar Y$. The crucial point is to use the
connection of the spin-1 of the initial state $J/\psi$ with the FS
with two spin-1/2. The crucial point is: the goal is to find
$\langle A^{(X)}_{\rm CP}\rangle \neq 0$ with{\em out} directly
measuring polarization of $Y/\bar Y$ \& $X/\bar X$ baryons and their
correlations due to the very narrow resonance $J/\psi$ in their
production.

One goes to measure correlations between the pair of FS baryons \&
pions. In the rest frame of $J/\psi$ one can define $C_T = (\vec p_X
\times \vec p_{\pi})\cdot \vec p_{\bar X}$ \& conjugate transitions
$\bar C_T = (\vec p_{\bar X} \times \vec p_{\bar \pi})\cdot \vec
p_{X}$ and compare the event numbers ($N$) with positive and
negative values
\bea
\langle A_T  \rangle &=& \frac{N(C_T > 0) - N(C_T <0)}{N(C_T > 0)+ N(C_T <0)} \\
\langle \bar A_T  \rangle &=&  \frac{N(\bar C_T > 0) - N(\bar
C_T<0)}{N(\bar C_T > 0) + N(\bar C_T <0)} \; . \eea Thus \beq
\label{eq:mathcalA_T} {\cal A}_T =  \frac{1}{2} \left[\langle A_T
\rangle + \langle \bar A_T \rangle\right] =\langle A_T\rangle  \neq
0 \eeq are observable CP asymmetries based on CPT invariance. We
have taken a different convention for $\bar A_T$ compared to
Ref.~\cite{LHCbLambda_b}.

One can compare to its charge-conjugate channel to get rid of a fake
CP asymmetry induced by the FSI effect; however, the
charge-conjugate of the process $J/\psi \to \Lambda \bar \Lambda \to
[p\pi^-][\bar p \pi^+]$ considered here is itself; or it is an
untagged sample that cannot be distinguished by experiment. Such
situation is, to some extent, different from a measurement of
$D^0(\bar D^0) \to K\bar K\pi\pi/4\pi$.

The landscapes are quite different for $\Lambda \to p \pi^-$ \&
$\Sigma \to p \pi$ and even more with $\Xi \to \Lambda \pi$ with
many `roads' for transitions as listed: $J/\psi \to \bar \Lambda
\Lambda \to [\bar p \pi^+][p \pi^-]$;\; $J/\psi\to\bar \Sigma ^-
\Sigma^+ \to [\bar p \pi^0] [p\pi^0]$;\; $J/\psi\to\bar \Xi^0 \Xi^0
\to [\bar \Lambda \pi^0][\Lambda \pi^0]$; \; $J/\psi\to\bar \Xi^+
\Xi^- \to [\bar \Lambda \pi^+][\Lambda \pi^-]$ \footnote{$M(\Lambda)
\simeq 1116$ MeV; $M(\Sigma^+) \simeq 1189$ MeV;  $M(\Xi^0) \simeq
1315$ MeV; $M(\Xi^-) \simeq 1322$ MeV.}.
\begin{itemize}
\item
One can calibrate those transitions
with $J/\psi \to \Delta (1232) \bar \Delta (1232)$, where CPV cannot appear there.

\item
That is not the end of the impact of strong resonances of $\Delta (1232)$ with the width
$\sim 117$ MeV and $N(1440)$ with the width $250 - 450$ MeV.
They will affect the lessons we learn from future data about possible CP violations in the transitions
of strange baryons.

\item
The item of "duality" between the worlds of quarks vs. hadrons is very subtle, in particular close to thresholds of resonances.

\end{itemize}
\begin{center}
\begin{table}[htbp]
\renewcommand{\arraystretch}{1.2}
\begin{center}
\begin{tabular*}{\linewidth}{@{\extracolsep{\fill}}ccc}
\hline \hline Channel & \# of events & Sensitivity on $\cal{A}_T$\\
\hline $J/\psi\to\Lambda\bar\Lambda\to [p\pi^-] [\bar p\pi^+]$ &
$2.6\times
10^6$ & 0.06\%\\
$J/\psi\to\Sigma^+ \bar \Sigma^- \to [p\pi^0] [\bar p\pi^0]$ & $2.5\times
10^6$ & 0.06\%\\
$J/\psi\to\Xi^-\bar\Xi^+\to [\Lambda\pi^-] [\bar \Lambda\pi^+]$
&$1.1\times 10^6$ &0.1\%\\
$J/\psi\to\Xi^0 \bar\Xi^0\to [\Lambda\pi^0] [\bar \Lambda\pi^0]$
&$1.6\times 10^6$ &0.08\%\\
 \hline\hline
\end{tabular*}
\caption{The  number of reconstructed events after considering the decay branching fractions, tracking, particle identifications.
The sensitivity is estimated without considering the possible background dilutions,
which should be small at the BESIII experiment.
Estimations are based on the $10^{10}$ $J/\psi$ data which will be collected by BES collaboration at 2018
(\& the branching fractions from PDG2016). Systematic uncertainties are expected to be at the same order as the
statistical one shown in the table.}
\label{EventNumber}
\end{center}
\end{table}
\end{center}
Here we estimate the sensitivities for measuring such observables
using the collected $J/\psi$ sample. In 2018 $10^{10} J/\psi$ events
will be accumulated by the BESIII experiment \cite{Li:2016tlt}. The
detection efficiency for pion, proton, kaon with momentum larger
than 100 MeV can reach 98\%. As for particle identification (PID),
the pion, kaon and proton can be distinguished with $3\sigma$ (three
standard deviation) below momentum of 1.0 GeV. Considering the
branching fractions of $J/\psi\to\Lambda\bar\Lambda\to [p\pi^-]
[\bar p\pi^+]$ and $J/\psi\to\Xi^-\bar\Xi^+\to [\Lambda\pi^-] [\bar
\Lambda\pi^+]$ \cite{Olive:2016xmw} etc., we can estimate the
expected number of events and the further sensitivities, see Table
\ref{EventNumber}. Probing such large data with re-fined tools will
give us information rich about the underlying dynamics.

\subsection{Going beyond first steps}
\label{BEYOND}

Measuring ${\cal A}_T$ is the first step to probe CP asymmetries.
There are possible options for intermediate steps like: \bea
\mathcal A_T (d) &=& \frac{N(C_T > |d|) - N(C_T < - |d|)}{N(C_T >
|d|) + N(C_T <-|d|)}
 \eea
It is a very promising way to go beyond ${\cal A}_T$ in
Eq.~\eqref{eq:mathcalA_T}. By the end of 2018 we can expect that
BESIII can probe CPV in the decays of strange baryons on the level
of $10^{-4}$ sensitivity, see in Tab.~\ref{EventNumber} (neglecting
the systematic errors).

The above method can be also applied to
$J/\psi\to\Lambda\bar\Lambda\pi\pi$ to probe the CPV in $\Xi$ decay.
Interestingly, for the case of $\Xi$ baryon the CPV can be also
probed by {\em polarized} $\Xi$ thanks to the decay chain $\Xi^0\to
\Lambda \pi\to (p\pi)\pi$ where the $CP$ violating observable can be
related to the helicity amplitudes, see a similar proposal in
Ref.~\cite{LambdaC} for $\Lambda_c$ decay. Such CP violating signal
can be extracted by performing an angular analysis which is again
accessible in the BESIII experiment due to the large $J/\psi$ data.
However, for $\Lambda$ decays the interference of the helicity
amplitude is absent in the angular observable \cite{BES1}, which
handicaps accessing $CP$ as the same way in polarized $\Xi$ decay,
i.e., measuring the interference of helicity amplitudes.

\subsection{$J/\psi \to \bar \Lambda \Lambda \to [\bar p \pi^+][p\pi^-]$}
\label{LAMBDA}

The DM2 Collaboration had measured CP invariance (and Quantum Mechanics) in
$e^+e^- \to J/\psi \to \bar \Lambda \Lambda \to [\bar p \pi^+][p\pi^-]$ in 1988
without polarized $\Lambda$ \& $\bar \Lambda$; their result $A_{\rm CP}=0.01 \pm 0.10$ \cite{DM2}.
It is the first step in an important direction.

Now we describe the situation 30 years later, say the potential what
BES can achieve by 2018. We use the Jacob-Wick helicity formalism
\cite{Jacob}, as one can see in the Fig.3 of the Ref.\cite{BES1}; it
was also applied in the Refs.\cite{KL,CPC}:
\begin{itemize}
\item
The $J/\psi$ rest frame is along the $\Lambda$ out-going direction, and the
solid angle $\Omega_0(\theta,\phi)$ is between the incoming $e^+$ \& out-going $\Lambda$.

\item
For $\Lambda \to p \pi^-$ the solid angle of the `daughter' particle $\Omega_1(\theta_1, \phi_1)$
is referred to the $\Lambda$ rest frame (although as out-going direction);
likewise for $\bar \Lambda \to \bar p \pi^+$.

\end{itemize}
We describe the angular distribution for this process following Ref.\cite{PingRG}:
\begin{eqnarray}
\label{eq:angular} \frac{d\Gamma}{d\Omega} &\propto&
(1-\alpha)\sin^2\theta \cdot
 \Big[1+\alpha_\Lambda\alpha_{\Lambar}\big(\cos\theta_1\cos\bar \theta_1 \nl
  && +\sin\theta_1\sin\bar \theta_1 \cos (\phi_1+\bar \phi_1 )\big)\Big] \nl
  && -(1+\alpha)(1+\cos^2\theta ) \left(\alpha_\Lambda\alpha_{\Lambar}\cos\theta_1
\cos\bar \theta_1-1\right) \, ,\nl \label{CPV1}
\end{eqnarray}
where $d\Omega \equiv d\Omega_0 d\Omega_1 d \bar\Omega_1$ and:
\begin{itemize}
\item
$\alpha$ is the angular distribution parameter for $\Lambda$;
\item
$\alpha_{\Lambda} [\alpha_{\bar \Lambda}]$ is the $\Lambda [\bar \Lambda]$ decay parameter;
\item
these data depend only on the product of $\alpha_{\Lambda}\alpha_{\bar \Lambda}$, see Eq.(\ref{CPV1}).

\end{itemize}
By fitting Eq.~(\ref{eq:angular}) to the data, one can determine
$\alpha_{\Jpsi}$ and $\alpha_{\Lambda}\alpha_{\Lambar}$.  One can
make a replacement: \beq \alpha_\Lambda\alpha_{\Lambar} \equiv
\frac{A-1}{A+1}\alpha_{\Lambda}^2  \; , \eeq where $A$ describes a
CP asymmetry observable.

As said before published 2010 data from the BES collaboration show
\cite{BES1} $\alpha^{(\bar p)}_{\bar\Lambda} = - 0.755 \pm 0.083 \pm
0.063$ based on previously measured $\alpha^{(p)}_{\Lambda} = 0.642
\pm 0.013$. Their non-zero numbers show the impact of re-scattering;
it is large, which is not surprising. We also note that
Eq.~(\ref{eq:angular}) is derived from considering only spin
projection $J_z=\pm1$ for $J/\psi$, which is a consequence of QED
process: $e^+e^-\to\gamma \to c\bar c\,\, (J/\psi)$
\cite{angular_cookbook}. Thus only the product term
$\alpha_{\Lambda}\alpha_{\Lambar}$ can be measured \footnote{In a
$p\bar p$ machine, the term $\alpha_{\Lambda}$ and
$\alpha_{\Lambar}$ can be separated alone
\cite{Donoghue_ppbar_Lambda}.}.

Present limit on direct CP asymmetry is around a few percent; hardly even ND cannot produce effects close to 1\% here.
To reach the ${\cal O}(0.1 \%)$ would be a sizable achievement about $\langle A^{( p)}_{\rm CP}\rangle$
based on $\alpha^{(p)}_{\Lambda} \sim \alpha^{(\bar p)}_{\bar \Lambda} \sim 0.64$.
Measuring semi-local asymmetry would give us new lessons about non-perturbative QCD.

It has been said very recently in the Ref.~\cite{Kupsc} that the
Ref.~\cite{BES1} missed some contributions for {\em on}-shell
$J/\psi \to \bar \Lambda \Lambda \to \bar p \pi^+ p \pi^-$; so far
we are not convinced. Of course, experimental data from the BESIII
are needed in order to test it.

Even now BESIII has many more data, and by 2018 we will have about
$2.6 \times 10^6$  events for $J/\psi \to \bar \Lambda \Lambda \to
[\bar p \pi^+] [p \pi^-]$ decay chain after considering the
efficiencies for tracking, particle identification and $\Lambda$
pair reconstruction~\cite{Li:2016tlt}. The sensitivity might reach
$6 \times 10^{-4}$ by 2018 as listed in Table \ref{EventNumber}. Now
the landscape is different with some hope to find CP asymmetry here,
and also for $J/\psi \to \bar \Sigma \Sigma$ that will be discussed
below.

General items: (a) Except $n - \bar n$ (or maybe, maybe even
$\Lambda - \bar \Lambda$) oscillations \cite{KangOsc} one probes
only direct CPV with baryons. (b) It is well-known that the impact
(local) penguin operators are crucial for $\Delta S=1$ transitions
for strange mesons, in particular about non-zero value for
$\epsilon^{\prime}$. What about decays of strange baryons? The
transitions of a pair of strange baryons are not far from
thresholds; thus one has to think about the item of "duality"
between the worlds of hadrons and quarks \& gluons. We come back
below.

To probe CP asymmetries with accuracy BESIII can calibrate with transitions where CP asymmetries cannot happen. We have two examples with
broad resonances, where the situations are `complex': $J/\psi \to \bar N(1440) N(1440)$ with $\Gamma_{N(1440)} \sim (250 - 450)$ MeV carrying isospin $I = 1/2$
and $J/\psi \to \bar \Delta (1232) \Delta (1223)$ with $\Gamma _{\Delta (1232)} \sim$ 117 MeV carrying isospin $I=3/2$. The BES experiment has measured the background very well
and continue to do it. The total background is very small, which provides good opportunity for a clean probe of CPV signal.

From the knowledge of the previous BES measurement \cite{BES1} (the
main background channels are also listed in this reference) and the
ongoing one \cite{BESpre}, one conservatively estimates  number of
the combinatorial background events is roughly $10^{-3}$ of the
signal events, a very small fraction such that the CPV can be
cleanly probed. That is one of the strengths of the BESIII
collaboration. We can show this more transparently with an example.
Till 2018 one expects $2.6\times10^6$ signal events, and assuming
the CPV is on the level of $10^{-4}$ one has $\Delta N=N_+
-N_-=260$. The background can induce $\Delta
N_{\textrm{bkg}}=\sqrt{2.6\times 10^3}\approx 51$. This illustrates
that the nonzero $A_T$ will really indicate the observation of a
$CP$ asymmetry. On the other hand, if the CPV $\mathcal{A}_T$ is
below the level of $10^{-5}$, the impact of the background is
important.

\subsection{$J/\psi \to \bar \Sigma^- \Sigma^+ \to [\bar p\pi^0][p\pi^0]$}
\label{SIGMA}

Above we have talked about $J/\psi \to \bar \Delta (1232) \Delta
(1232)$ as background for $J/\psi \to \bar \Lambda \Lambda$. We had
said before it is very important to analyze pairs of baryons.
However the situation here is even more `complex', as we have discussed just above:
\begin{itemize}
\item
$\Sigma^+$ carries isospin $(I, I_3)=(1, +1)$, while $\Lambda$
isospin zero.

\item
Looking on straightforward diagrams we get $J/\psi \to \bar \Sigma^- \Sigma^+ \to [\bar p \pi^0][p\pi^0]$
vs. $J/\psi \to \bar \Delta (1232) \Delta (1232) \to \bar p \, \pi\, p \,\pi$ as a background.

\item
However, we have to go beyond that as you can see by comparing
$M(\Sigma^+) \simeq$ 1189 MeV vs. $M(\Delta (1232))\simeq$ 1232 MeV
with the width of $\Delta (1232) \simeq$ 117 MeV. {\em Off}-shell
intermediate amplitudes sizably affect total amplitudes and also
measurable CP asymmetries in $[\bar p\pi^0][p\pi^0]$ final states.

To say it with different words, but with the same meaning: there would be a
sizable overlaps between the waves of $\bar \Sigma$/$\Sigma$ and $\bar \Delta (1232)$/$\Delta (1232)$ due to $\Gamma (\Delta (1232))$. Therefore we cannot ignore that.

\item
Therefore we use two different diagrams for transitions:
"$\Rightarrow$" describes the amplitudes due to  QCD($\times$QED) that conserve P \& C
symmetries; "$\to$" includes $SU(2)_L$ with weak dynamics including sources of P, C
and CP asymmetries. The direct `road' is
\beq
J/\psi \Rightarrow \bar \Sigma^- \Sigma^+  \to [\bar p\pi^0][p\pi^0] \; ,
\eeq
while also a somewhat indirect one can happen due to off-shell intermediate amplitudes:
\beq
J/\psi \Rightarrow "\bar \Delta (1232) \Delta (1232)" \Rightarrow
\bar \Sigma^- \Sigma^+  \to [\bar p\pi^0][p\pi^0] \; .
\eeq
As said before, the impact of re-scattering is `complex',
as one can see from comparing BR$(\Sigma^+ \to p\pi^0)\simeq 0.52$ \&
BR$(\Sigma^+ \to n\pi^+)\simeq 0.48$ vs. BR$(\Lambda \to p\pi^-)\simeq 0.64$ \&
BR$(\Lambda \to n\pi^0)\simeq 0.36$.

\item
There is another possible amplitude that is even more `complex', in
particular with \beq J/\psi \Rightarrow \bar \Sigma^- \Sigma^+ \to
"\bar \Delta (1232) \Delta (1232)" \Rightarrow [\bar p\pi^0][p\pi^0]
\; . \eeq

\end{itemize}
We `paint' off-line exchanges of kaons or new dynamics.

The data produced by 2018 will be analyzed by the BESIII
collaboration with the best tools. It is possible that off-line
resonances like $\bar \Delta (1232) \Delta (1232)$ might have an
impact on CP asymmetries more on $J/\psi \to \bar \Sigma^- \Sigma^+
\to [\bar p \pi^0][p \pi^0]$ than in $J/\psi \to \bar \Lambda
\Lambda \to [\bar p \pi^+][p \pi^-]$, since off-shell $\bar \Delta
(1232) \Delta (1232)$ are closer to the on-shell $\bar \Sigma^-
\Sigma^+$; the second vertex technique will help much to distinguish
them from the background from $\bar \Delta (1232)\Delta (1232)$.

Using $10^{10} J/\psi$ data that will be collected by the BESIII until
end of 2018, we expect $2.5\times 10^6$ signal events, with
sensitivity at $0.06\%$, see Table~\ref{EventNumber}.

\subsection{$J/\psi \to \bar \Xi \Xi \to [\bar \Lambda \pi][ \Lambda \pi]$}
\label{XiXiLambda}

It is also interesting to probe $CP$ violation by using the decays
$J/\psi \to \bar \Xi^+ \Xi^-$ and  $\bar \Xi^0 \Xi^0$. One can
reconstruct both $\Xi$s in the $\Xi \to \Lambda \pi$ mode.  For the
neutral channel $\Xi \to \Lambda \pi^0$, the $\pi^0$ in $\bar \Xi^0
$ decay can be easily separated from that in $\Xi^0 $ decay without
ambiguity, since both $\Xi$s are back to back and strongly boosted
in the rest frame of $J/\psi$, namely, the $\Xi^0 \to \Lambda \pi^0$
is reconstructed in the opposite decay hemisphere against the decay
hemisphere for the $\bar \Xi \to \bar \Lambda \pi^0$. This situation
will be similar as in Sec.~\ref{SIGMA} for $J/\psi \to \bar \Sigma^-
\Sigma^+ \to [\bar p\pi^0][p\pi^0]$. By 2018 we will have data for
$J/\psi \to \bar \Xi^+ \Xi^- \to \bar \Lambda \pi^+ \Lambda \pi^-$
and $\bar \Xi^0 \Xi^0 \to \bar \Lambda \pi^0 \Lambda \pi^0$ with the
numbers of $1.1\times 10^6$ and $1.6\times10^6$~\cite{Li:2016tlt},
respectively.  Thus, the reaches for the triple-product asymmetry
are about $10 \times 10^{-4}$ and $8 \times 10^{-4}$ for the charged
and neutral $\Xi$, respectively.

Once non-zero values for CP asymmetries are found, one can attract
another challenge: is it source of the transition of $\Xi \to
\Lambda \pi$ or $\Lambda \to p \pi^-$ -- or the interferences
between them?

\subsection{$J/\psi \to \Lambda \bar X$ vs. $J/\psi \to \bar \Lambda X$ }
\label{YY}

One can also compare the transitions $J/\psi \to \Lambda \bar X$ vs. $J/\psi \to \bar \Lambda X$, in particular
$J/\psi \to \Lambda \bar p K^+$ vs. $J/\psi \to \bar \Lambda  p K^-$. 2016 data tell us:
BR$(J/\psi \to \Lambda \bar p K^+/\bar \Lambda p K^-) = (0.89 \pm 0.16)\cdot 10^{-3}$.
By 2018 the expected number of events is about
$9 \times 10^6$ which has sensitivity for CP asymmetry of $2.4\times 10^{-4}$ by comparing the partial widths between
$J/\psi \to \Lambda \bar X$ and  $J/\psi \to \bar \Lambda X$ decays. The systematic uncertainty are
expected larger than in the previous cases discussed above, but not by an order of ten.

\subsection{Summary from the experimental side}
\label{FIRSTSUM}

We learn about non-perturbative QCD with a novel situation; it is not trivial how much to apply them
and where. Just below we give comments about available theoretical tools.
The hope is to find signs of CP asymmetries in the decays of strange baryons; therefore the goal is not go for
accuracy. We `paint' the landscape to find CP asymmetries. Once our community has found
non-zero value somewhere, one can discuss the correlations with other situations.

\section{Comments about tools for CP asymmetries}
\label{TT}

The realistic goal is to get new lessons about non-perturbative QCD; one can {\em hope} to find CP asymmetries in the decays of strange
baryons; one can also dream to find the impact of ND \cite{CPBOOK}.
When one goes after accuracy, one needs
truly consistent parametrization of the CKM matrix \cite{AHN}. However, that is not the goal now; therefore one can use the well-known Wolfenstein's parametrization.

We have {\em local} operators for $\Delta S =1$ amplitudes. One
describes the scattering of left-handed quarks $s_L + u_L \to d_L +
u_l$ with re-fined tree amplitudes and {\em local} penguin
operators. Challenges come from true strong dynamics in different
ways. In particular $\mu \sim 1.0$ GeV describe the "fuzzy"
boundaries between perturbative and non-perturbative QCD. On the
other side the landscape is also complex, since the baryons carry
spin-1/2; therefore there are more observables.  To use different
words, but the same substance: the amplitudes can be described with
S- or P-waves.

There is another challenge, namely to connect quark and hadronic
amplitudes. Namely, the item of "duality" is well known, and it is
not just another assumption based on true quantum field theory.
However, it does not work well, when one has to deal with thresholds
that are important in these on-shell transitions $J/\psi \to \bar
\Lambda \Lambda$ and $J/\psi \to \bar \Sigma^- \Sigma^+$; in these
cases we have broad resonances like $\Delta (1232)$ and $N(1440)$.

\section{About the future}
\label{SUM}

As said above a realistic goal is to measure non-leptonic decays of
$\Lambda$, $\Sigma^+$ and $\Xi$ with more data to learn new  lessons
about non-perturbative QCD. One can `hope' to find CP asymmetries in
BESIII data by the end of 2018 and `dream' to find the impact of ND.
When it is {\em not} enough to work on data \& their analyses where
one `hope' or even `dream 'to learn from that, then -- in our view
-- one is in wrong business.

Of course, we need much more data, but also powerful analyses to
reach even the realistic goal. Here we have listed the directions,
where more data should improve our understanding of fundamental
dynamics, see the Tab.~\ref{EventNumber}. In a future super
tau-charm factory
\cite{superTauCharm1,superTauCharm2,superTauCharm3}, one will have
an unprecedentedly high peak luminosity of $10^{35}$
cm$^{-2}$s$^{-1}$, with $10^{12} J/\psi$ data samples, 100 times as
large as the ongoing BEPCII/BESIII, which will result in a
decreasing of the (statistical) uncertainty by 10 times.

We do not give predictions here for CP asymmetries.  Our goal here
is to point out the `road(s)', where our community can learn more in
the future. We need more thinking in general and analyze
correlations in different transitions.

Analyzing $e^+e^- \to \bar \Lambda_c^-\Lambda_c^+$ can also give us
novel lessons about non-perturbative QCD and even to compare
$e^+e^- \to \bar \Lambda_c^- \Lambda_c^+\to \Lambda +\bar X$ vs.
$e^+e^- \to \bar \Lambda_c^- \Lambda_c^+\to\bar \Lambda + X$.

We add a comment that one first sees it as a technical one: applying
"dispersion relations" has a long history about nuclear physics,
hadrodynamics and High-energy physics (HEP) and also other branches
of physics. They are based on central statements of quantum field
theory, while their results depend on low energy data with some
`judgment'. If the 2018 data we have discussed above exist, there is
a good chance to convince members of our community to work on that.
There are two jobs to do applying dispersion relations: (i) As an
input for them one needs more data from $N\pi$ collisions at low
energies. (ii) Non-trivial thinking is needed where \& how to apply
them for good reasons.

Finally for the future our community could have a novel competition
between BESIII, LHCb and theorists -- actually, parts of a team.

\vspace{0.5cm}

{\bf Acknowledgments:} XWK thanks for the careful reading,
encouragement and interesting discussion from Hai-Yang Cheng, and
also useful discussion with Xinxin Ma. This work is supported by the
National Science Foundation under the grant number PHY-1520966; in
part by the National Natural Science Foundation of China under
contracts Nos. 11335009, 11125525, Joint Large-Scale Scientific
Facility Funds of the NSFC and CAS under Contracts No. U1532257; the
National Key Basic  Research Program of China under Contract No.
2015CB856700 and also supported by Key Research Program of Frontier
Sciences, CAS, Grant No. QYZDJ-SSW-SLH003. XWK's work is also
supported by the Ministry of Science and Technology of R.O.C. under
Grant No. 104-2112-M-001-022.

\vspace{4mm}


\end{document}